\documentclass[prl,twocolumn,superscriptaddress,showpacs]{revtex4}
\usepackage{amsmath,amssymb,graphicx}

\begin{document}
\title{Comment on "Experimental Demonstration of the Time Reversal Aharonov-Casher Effect"}

\author{Y. Lyanda-Geller}
 \affiliation{Department of Physics and Birck Nanotechnology Center, Purdue University, West Lafayette IN 47907, USA}

\author{I. A. Shelykh}
 \affiliation{International Center for Condensed Matter Physics, 70904-970 Brasilia-DF, Brazil
 and St. Petersburg State Polytechnical University, 195251, St. Petersburg, Russia}

\author{N.T. Bagraev}
 \affiliation{A.F.Ioffe Physico-Technical Institute of RAS, 194021, St. Petersburg, Russia}

\author{N.G. Galkin}
 \affiliation{A.F.Ioffe Physico-Technical Institute of RAS, 194021, St. Petersburg, Russia}

\date{\today}


\maketitle

In a recent Letter \cite{Nitta}, Bergsten et al. have studied the resistance oscillations with gate voltage $V_g$ and
magnetic field $B$ in arrays of semiconductor rings, and interpreted the oscillatory $B$-dependence 
as Altshuler-Aronov-Spivak oscillations and oscillatory $V_g$-dependence as the time reversal Aharonov-Casher (AC) effect. 
This comment shows (i) that authors \cite{Nitta} incorrectly identified AAS effect as a source of resistance
oscillations with $B$, (ii) that spin relaxation in [1] is strong enough to
destroy oscillatory effects of spin origin, e.g., AC effect, and (iii) the oscillations in \cite{Nitta} are 
caused by changes in carrier density and the Fermi energy by gate, and are unrelated to
spin.

AAS effect is the $h/2e$ oscillations of conductance with $B$ in disordered diffusive rings. 
Oscillations occur because the intereference of the two electron trajectories passing the whole ring clock- and counterclockwise 
survives disorder averaging in conditions of diffusive
regime $l \ll L_{\phi},L$, where $l$ is the mean free path, $L_{\phi}$ is the phase breaking length, and $L$ is the
circumference of the ring. 

The mean free path in samples \cite{Nitta} is $l\sim1.5{-}2\,\mu{m}$. From the ratio of h/2e and h/4e signal
amplitudes \cite{Nitta}, $L_{\phi}$ is between 2.8 and 3.5 $\mu{m}$. (Note that $h/2e$ signal is due to interference
of clockwise and counterclockwise paths, with magnitude defined by $\exp(-2L/L_{\phi})$, and $h/4e$
oscillations are due to interference of paths going twice clockwise and twice counterclockwise, defined by $\exp(-4L/L_{\phi})$. 
The calculation of $L_{\phi}$  in \cite{Nitta} misses a factor of two.). 
Thus, samples \cite{Nitta}  are not in diffusive regime relevant to AAS oscillations, but are in the quasi-ballistic regime
$l \lesssim L$.  Then $h/2e$ oscillations are defined not only by interference of time-reversed paths, but also e.g.,
by the interference of the amplitude of propagation through the right arm clockwise and the amplitude of propagation via the  
 three-segment path: the
left arm, the right arm (counterclockwise) and again through the left arm. With all interference processes
included, $h/2e$ oscillations depend on the Fermi wave-vector and $n_s$ \cite{Buttiker},\cite{Aronov}. Averaging over
few resistance curves does not eliminate contributions of of non time-reversed processes (certainly not
beyond $0.3{\%}$ of the overall signal for oscillations in \cite{Nitta}). Their importance is
missed in \cite{Nitta} and is crucial.

    (ii) Another mistake in \cite{Nitta} is the neglect of spin
relaxation. For the spin-orbit constant $\alpha=5\,\mathrm{peV}{\cdot}\mathrm{m}$, the parameter $\alpha ml\sim2.5$ (m is the effective  mass), and spin simply
flips due to a single scattering event. The spin-flip length $L_S=l< L$. Thus, oscillations of spin origin are rather
unlikely in \cite{Nitta}. The closest to \cite{Nitta} feasible setting requires
ballistic regime $l \gg L$ \cite{Aronov}, which requires mobility an order of magnitude higher. Note
that $L_{\phi}>L_S$ and oscillations with $B$ originating from charge coherence are plausible to observe.

    (iii) The key to understanding the $h/2e$ oscillations with $V_g$ in \cite{Nitta} is it's Fig .4.
It can be seen clearly that resistance oscillations are present only when $n_s$ changes with $V_g$, and are not present when $n_s$ saturates.
Therefore the reason for the observed oscillations is the variation of the $n_s$. Oscillations of spin origin, particularly the AC effect,
must persist when $n_s$ is constant, while $\alpha$ varies with $V_g$. No such evidence is present in \cite{Nitta}.

    The origin of oscillations with $n_s$ is the contribution to $h/2e$ signal from interference of non-time reversed paths.
These are independent of $L_S$, and are governed by $L_{\phi}>L_S$. That makes this effect dominant over any spin oscillations.
With the account of the role of contacts connecting the ring and the leads \cite{Buttiker},\cite{Aronov}, in the absence of spin-orbit interactions
and for strong coupling of leads and rings, the conductance of the single ring is
\begin{equation}
\label{conductance}
G=\frac{2e^2}{h}\left[1-\left|\frac{1-\cos\left(\pi\Phi/\Phi_0\right)}{1-e^{ik_FL}\cos^2\left(\pi\Phi/\Phi_0\right)}\right|^2\right]
\end{equation}

\begin{figure}[h]
\includegraphics[width=1.0\linewidth]{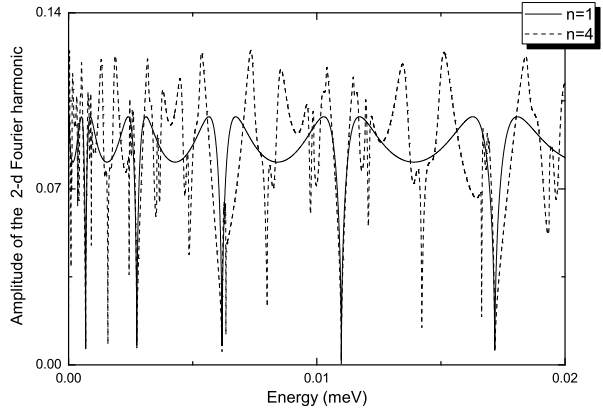}
\caption{\label{fig1} The amplitudes of the second harmonics in a single ring (solid curve) and four consequently connected rings (dashed curve).
The spin-orbit interaction is absent.}
\end{figure}

We note that disregard of transmission and reflection from contacts is yet another critical omission in \cite{Nitta}, whose equation for conductance is incorrrect in ballistic/quasi-ballistic regime. (It is also incorrect for AAS and AC effect in diffusive regime).
The second harmonics in (\ref{conductance}) depends on $k_F$ and $n_s$ in an oscillatory manner, 
leading to oscillations of conductance with $V_g$.
The system of the n interconnected rings can be described similarly to the setting in 
\cite{Shelykh}. 
On Fig. 1, we show the dependence of the amplitude of the second harmonic on
$k_F$ for one and four rings. Conductance oscillates with electron density despite no spin effects are involved.
To summarize, conclusions of \cite{Nitta} on the observation of the AC effect are unfounded.

\end{document}